\documentclass{PoS}
\title{Quarkonium production as a function of charged-particle multiplicity in pp and p--Pb collisions measured by ALICE at the LHC}

\ShortTitle{Quarkonium vs. multiplicity}

\author{\speaker{Dhananjaya Thakur for the ALICE Collaboration}\\
{Discipline of Physics, School of Basic Sciences, Indian Institute of Technology Indore, Indore- 453552, INDIA}
   \newline     
   E-mail: \email{dhananjaya.thakur@cern.ch}}

\abstract{
Quarkonium production as a function of the charged-particle multiplicity could provide an insight into particle production processes at the partonic level in hadronic collisions. It is believed that multiple partonic interactions play an important role in particle production and affect both soft and hard processes. The study of correlations between quarkonia and charged-particle multiplicity may provide information about this. In this contribution, ALICE measurements of J$/\psi$ and $\Upsilon$ production as a function of charged-particle multiplicity are presented for \rm{pp} collisions at center-of-mass energies $\sqrt{s}$ = 5.02  and 13 TeV.  
 A similar measurement performed in p\textendash Pb collisions at $\sqrt{s_{\rm{NN}}}$ = 8.16 TeV  at both forward and backward rapidity is also discussed.
}
\FullConference{International Conference on Hard and Electromagnetic Probes of High-Energy Nuclear Collisions\\
		30 September - 5 October 2018\\
		Aix-Les-Bains, Savoie, France}

\begin{document}

\section{Introduction}
\label{intro}
In pp collisions at the LHC, where the center-of-mass energy is very high, the measurement of the inclusive production of charged hadrons and their correlation to quarkonia is an important tool to characterize  particle production. Quarkonium (bound state of charm and anti-charm quarks or beauty and anti-beauty quarks) production in \rm{pp} collisions allows us to test perturbative QCD calculations and is used as a reference for studies in heavy-ion collisions. The interplay between hard and soft QCD processes in hadronic collisions can be studied by measuring quarkonium production as a function of charged-particle multiplicity. This study is believed to bring information about the structure of the collision in terms of Multiple Partonic Interaction (MPI), in which several interactions occur in parallel in a single pp collision.  The role of MPI in heavy-flavor production has already been investigated by measuring the relative J$/\psi$ and D-meson yields as a function of charged-particle multiplicity in \rm{pp} collisions at $\sqrt{s}$ = 7 TeV~\cite{Adam:2015ota}. In both cases, an increasing trend has been observed. With the help of new data at $\sqrt{s}$ = 5.02 and 13 TeV, one can study even larger relative multiplicities and learn about the energy dependence. Similarly, the study of the multiplicity dependence of J$/\psi$ production in p--Pb collisions at $\sqrt{s_{\rm{NN}}}$ = 5.02 TeV is extended to $\sqrt{s_{\rm{NN}}}$ = 8.16 TeV, for which higher relative multiplicity can be reached, compared to previous measurements.

\section{Experimental setup and analysis procedure}
\label{sec:expt}
In ALICE~\cite{Aamodt:2008zz} , J$/\psi$ mesons are reconstructed via the  J$/\psi \rightarrow \rm{e}^{+}  \rm{e}^{-}$ and J$/\psi \rightarrow \rm{\mu}^{+}  \rm{\mu}^{-}$ decays using two different spectrometers. The central barrel detector covers the pseudo-rapidity range $|\eta| < 0.9$ and includes the Inner Tracking System (ITS) and Time Projection Chamber (TPC). It is used for the reconstruction of J$/\psi$ via the $\rm{e}^{+}  \rm{e}^{-}$ decay channel in the rapidity range $|y|< 0.9$. In contrast, J$/\psi$ are reconstructed in the rapidity range $-4.0 < y < -2.5$ via the $ \rm{\mu}^{+}  \rm{\mu}^{-}$ decay channel using the forward muon spectrometer. Two V0 detectors, V0A and V0C, consisting of two scintillator arrays located at $-3.7 < \eta < -1.7$ and at 2.8 $< \eta <$ 5.1 are used for triggering. The V0 detectors are also used as a high-multiplicity trigger. Tracklets, i.e. track segments reconstructed in the Silicon Pixel Detector (SPD) at $|\eta| < 1$, are used for the determination of charged-particle pseudo-rapidity density ($\rm{d}\it{N}_{\rm{ch}}/\rm{d}\eta$). Several cuts are applied to determine the accurate position of the $\it{z}$-coordinate of the collision vertex. The  $z$ positions of the collision vertex is restricted to the range $ |z_{\rm{vtx}}| < $ 10 cm, to have sufficient acceptance of the SPD for tracklets to be measured in the range $|\eta|<1$.
%In order to ensure a sufficient acceptance of the SPD for tracklets in the range $|\eta|<1$, the analysis is restricted $z$ position of the collision vertex in the range $ |z_{\rm{vtx}}| < $ 10 cm. 
%To account for the SPD acceptance and efficiency, a $z_{\rm{vtx}}$-dependent correction is applied using a data driven method~\cite{javier_thesis}, in which the correction factors to the measured %number of tracklets are randomized on an event-by-event basis using a Poissonian distribution, in order to preserve the integer nature of this quantity and to avoid artifacts when defining multiplicity %intervals. 
The actual charged-particle density in the $i^{\rm{th}}$ multiplicity interval is calculated as:
\begin{equation}
\scalebox{1.3}
{
$\frac{{\langle d\it{N}_{\rm{ch}}/d\eta \rangle }_{i}}{{\langle d\it{N}_{\rm{ch}}/d\eta \rangle }}  =  \frac{{f( \langle N_{\rm{trk}}^{\rm{corr}} \rangle }_{i})}{\langle  dN_{\rm{ch}}/d\eta \rangle }_{\scalebox{0.5}{\rm{INEL }> 0}}$,
}
\label{eq:inel_nch}
\end{equation} 
where~${\langle  N_{\rm{trk}}^{\rm{corr}} \rangle }_{i}$ is the mean number of corrected SPD tracklets in $i^{\rm{th}}$ multiplicity interval. The correlation function $f$ is calculated using Monte Carlo (MC) simulations. The measurement is performed for inelastic \rm{pp} collisions with at least one charged particle  in $|\eta|< 1$.
The relative J$/\psi$ yield in the $i^{\rm{th}}$ charged-particle multiplicity interval is calculated as:

\begin{equation}
%\begin{aligned}
\label{eq:jpsi_relativeyield}
\frac{\frac{\rm{d}\it{N}_{\rm{J}/\psi}^{\rm{~i}}}{\rm{d}y}}{\langle \frac{\rm{d}\it{N}_{\rm{J}/\psi}}{\rm{d}y}  \rangle} = \frac{\it{N}_{\rm{J}/\psi}^{~\it{i}}}{\it{N}_{\rm{J}/\psi}} \times \frac{\it{N}_{\rm{MB}}}{\it{N}_{\rm{MB}}^{\it{i}}} \times \epsilon,
%\end{aligned}
\end{equation} where ($N_{J/\psi}^{\it{i}}$, $N_{\rm{J/\psi}}$)  and   ($N_{\rm{MB}}^{i}$, $N_{\rm{MB}}$) are the number of J$/\psi$ and number of minimum bias events (V0-triggered events) in the $i^{\rm{th}}$ multiplicity interval and integrated  over all multiplicity intervals, respectively. The factor $\epsilon$ corrects for the efficiencies  concerning trigger selection, event cuts, pile up rejection and correction factor for INEL> 0 event selection.

For the analysis of p--Pb collisions, correction factors are applied on both the measured J$/\psi$ relative yields and the relative charged-particle density so that they correspond to Non-Single-Diffractive (NSD)~\cite{Adamova:2017uhu} events. Additionally, the J$/\psi$ mean transverse momentum ($\langle p_{\rm{T}}^{\rm{J}/\psi} \rangle$) is measured as a function of the charged-particle multiplicity by fitting the $\langle p_{\rm{T}} \rangle$ of unlike-sign muon pairs as a function of the dimuon invariant mass.

\section{Results}
\label{sec:result}

\begin{figure*} [h]
\begin{center}
%\begin{floatrow}
\includegraphics[width=6.2cm,height=4.3cm]{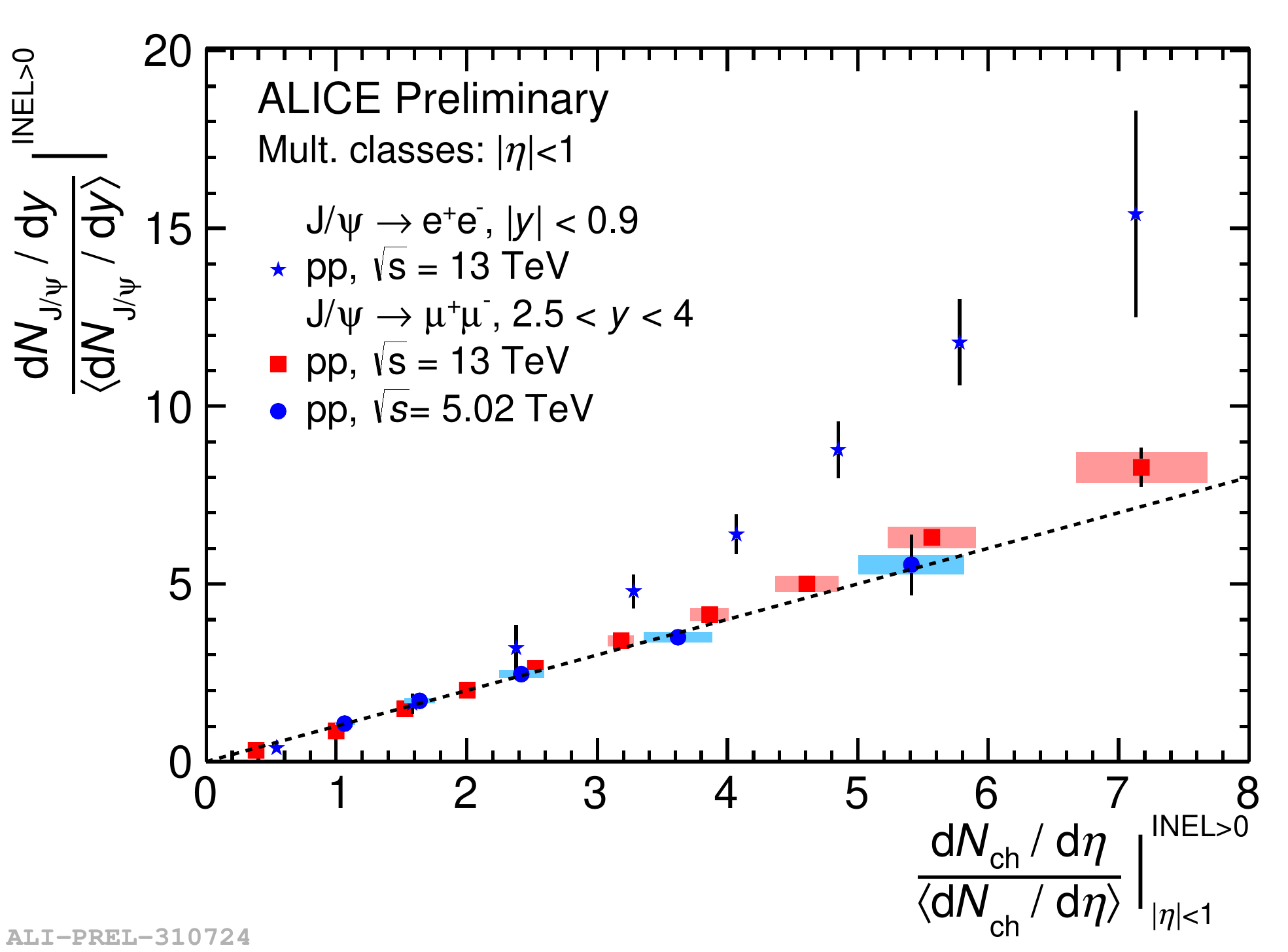}
\includegraphics[width=6.1cm,height=4.2cm]{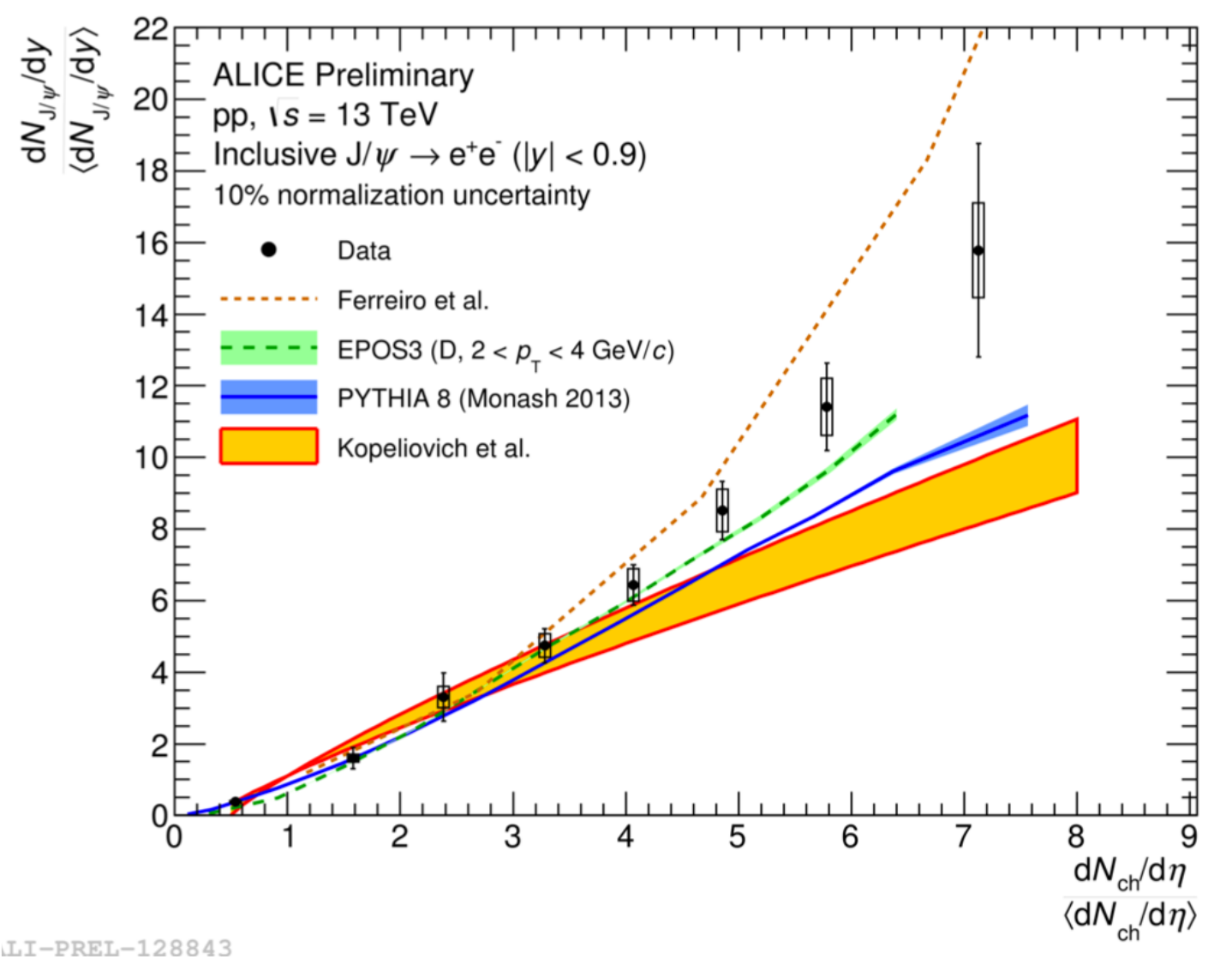}
\includegraphics[width=14.3pc]{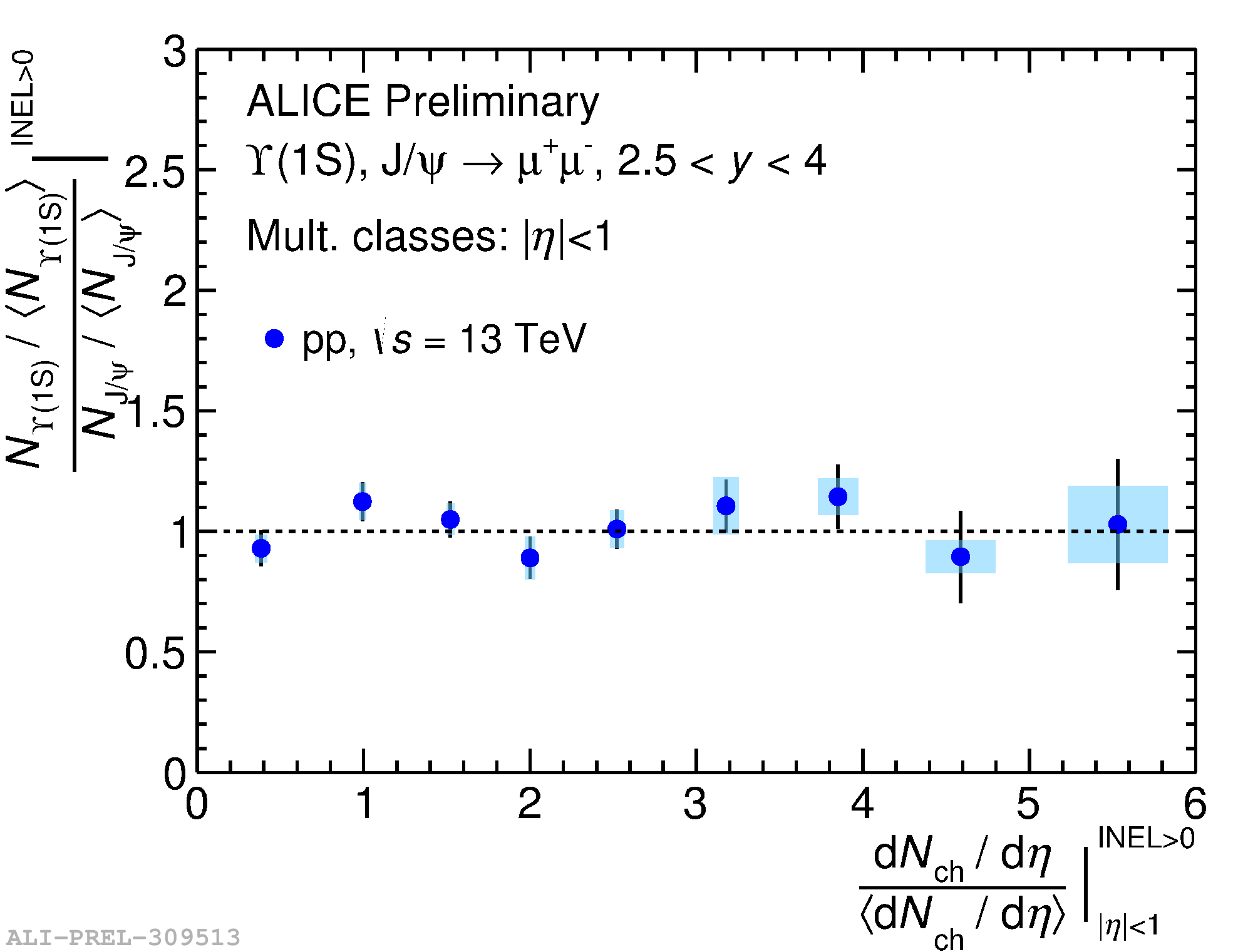}
\includegraphics[width=14.3pc]{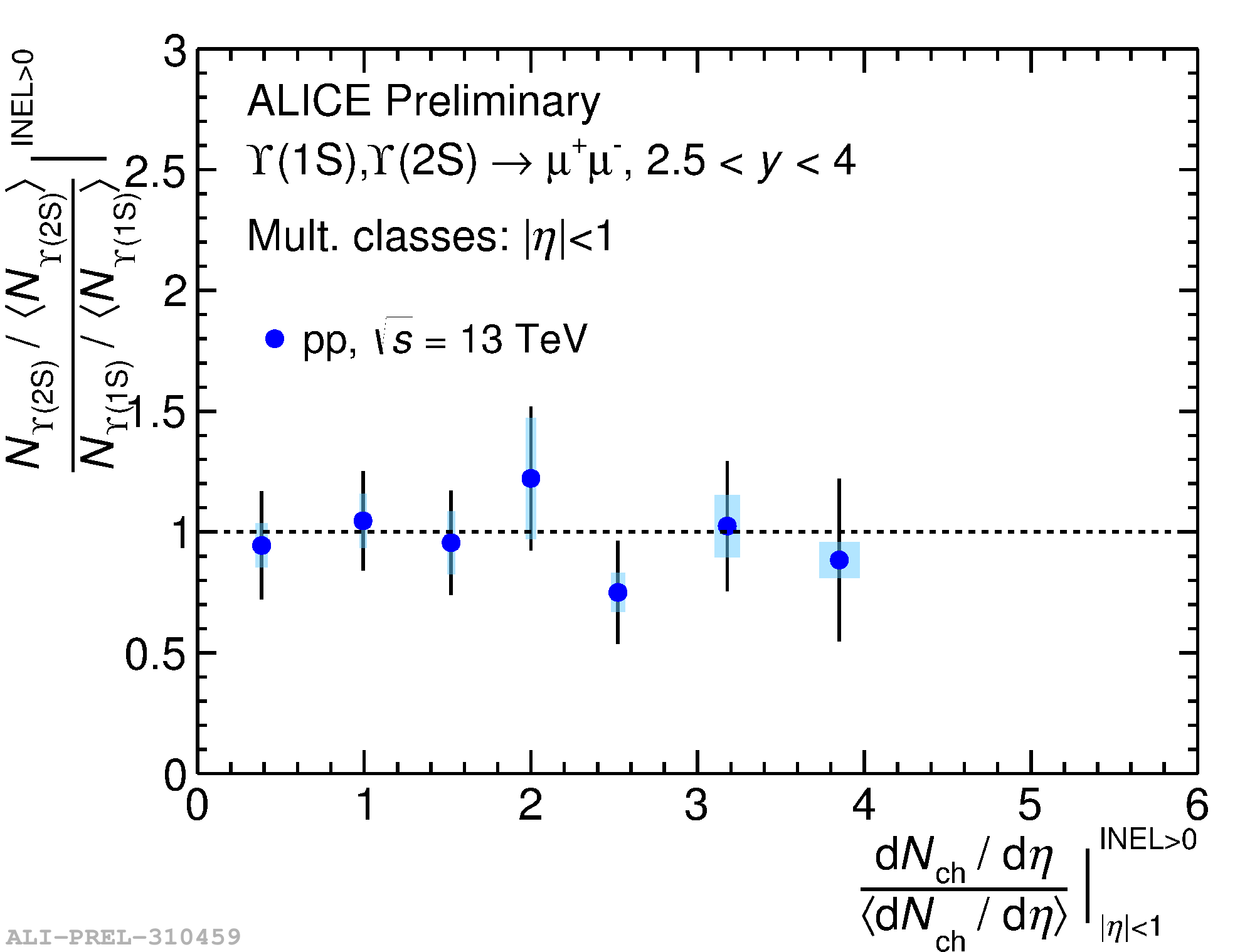}
%\end{floatrow}
\caption{Top left: relative J$/\psi$ yield as a function of the relative charged-particle density measured at mid and forward rapidity in pp collisions at $\sqrt{s} = 13$ TeV and at forward rapidity in pp collisions at $\sqrt{s} = 5.02$ TeV. The dotted line corresponds to a linear increase. Top right: model comparison for the mid-rapidity result at $\sqrt{s} =13$ TeV. Bottom panels: ratio of relative yield of $\Upsilon(1\rm{S})$ to J$/\psi$ (left) and $\Upsilon(2\rm{S})$ to $\Upsilon(1\rm{S})$ (right) as a function of multiplicity.}
\label{fig1}
\end{center}
\end{figure*}

The top left panel of Fig.~\ref{fig1} shows the $p_{\rm{T}}$-integrated relative J$/\psi$ yield at mid and forward rapidity as a function of charged-particle multiplicity in \rm{pp} collisions at $\sqrt{s} = $13 TeV, compared to that at $\sqrt{s} =$ 5.02 TeV.  In all cases the J$/\psi$ yield increases with increasing charged-particle multiplicity.  A similar trend has been observed for \rm{pp} collisions at $\sqrt{s} =$ 7 TeV and it is believed that this behavior is due to the role of MPI on the production of J$/\psi$~\cite{aliceplb,Thakur:2017kpv}. 
%For forward-rapidity measurements, the increase is approximately linear, whereas for the mid-rapidity measurement, the increase is faster than linear. 
It can also be observed that the relative J$/\psi$ yield increases as a function of relative charged-particle density faster than linear when the J$/\psi$ signal and the multiplicity are measured in the same rapidity window (mid-rapidity). On the contrary, with the introduction of a rapidity gap between the signal ($2.5 < y < 4.0$) and the charged-particle density estimator (|$\eta| < 1$), an approximately linear increase of the relative J$/\psi$ yield with increasing relative charged-particle density is observed. The different trends observed with the introduction of a rapidity gap might hint to autocorrelations or jet biases. One can also observe that the multiplicity dependence of the J$/\psi$ production at forward rapidity is the same in \rm{pp} collisions at $\sqrt{s}$ =13 TeV and 5 TeV, indicating that particle production is dominantly driven by the relative event activity independently of $\sqrt{s}$. 
%A similar linear increase, and absence of energy dependence has also been observed in the light flavor sector~\cite{Vislavicius:2017lfr} . 

In the top-right panel of Fig.~\ref{fig1}, the mid-rapidity measurement in pp collision at $\sqrt{s}$ = 13 TeV is compared to different theoretical models in which MPI is an important ingredient. They are: Ferreiro et al.~\cite{Ferreiro:2012fb}, EPOS3~\cite{Drescher:2000ha}, PYTHIA~8~\cite{Sjostrand:2007gs}~and~Kopeliovich et al.~\cite{Kopeliovich:2013yfa}. All the models qualitatively reproduce the trend and EPOS3 reproduces the behavior quantitatively well. EPOS3 includes MPI and hydrodynamical expansion of the system and describes the azimuthal correlation of D mesons with charged particles well~\cite{ALICE:2016clc}. The EPOS3 calculation shown in this work is for D mesons in a $p_{\rm{T}}$ range that is supposed to serve as good proxy of inclusive J$/\psi$ production. The good agreement of the EPOS3 model with the data might show that the energy density reached in pp collisions at the LHC might be high enough to be described by a hydrodynamical evolution, but other explanations are also possible. Additionally, the multiplicity dependence of J$/\psi$ production has also been studied for several J$/\psi$ $p_{\rm{T}}$ intervals. In that case, the observed trends are qualitatively well reproduced by PYTHIA 8 simulations.

To investigate the possible dependence of the observed increase on the quark content of the quarkonium state, and on its mass, a similar study of the relative $\Upsilon$(1\rm{S}) and $\Upsilon$(2\rm{S}) yields as a function of the relative charged-particle density has been performed in \rm{pp} collisions at $\sqrt{s}$ = 13 TeV. A linear increase of the relative $\Upsilon$(1S) and $\Upsilon$(2S) yields as a function of multiplicity is observed as well. To better quantify the similarities between the different particle species, we have measured the $\Upsilon(1\rm{S})$ to J$/\psi$ and the $\Upsilon(2\rm{S})$ to $\Upsilon(1\rm{S})$ relative yield ratios, as a function of multiplicity. It can be seen in the bottom panels of Fig.\ref{fig1} that these ratios are consistent with unity for all multiplicities. This study reveals that the multiplicity dependence is the same for the various quarkonium states despite their different masses and binding energies.

\begin{figure*} [h]
\begin{center}
\includegraphics[width=14.2pc]{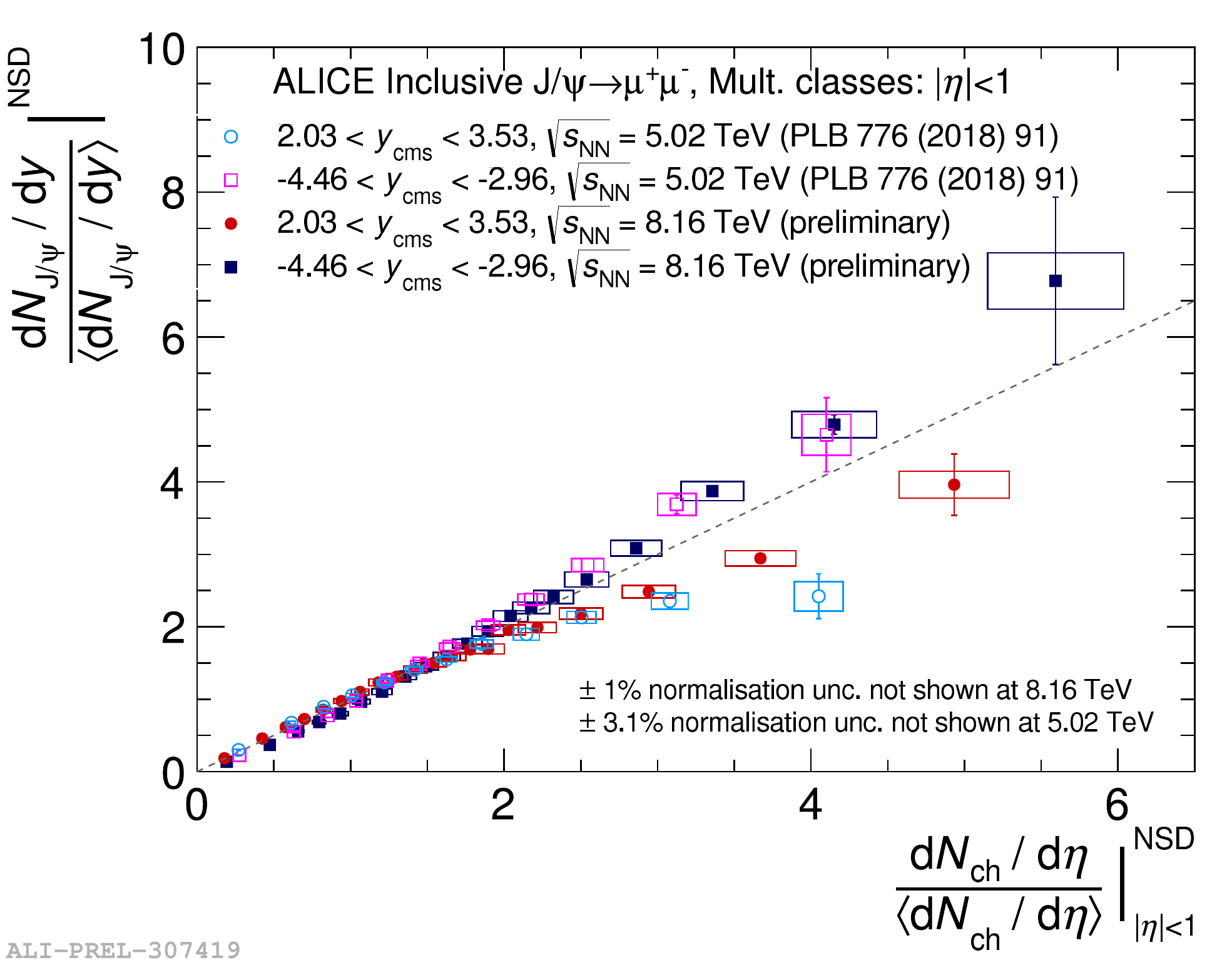}
\includegraphics[width=14.2pc]{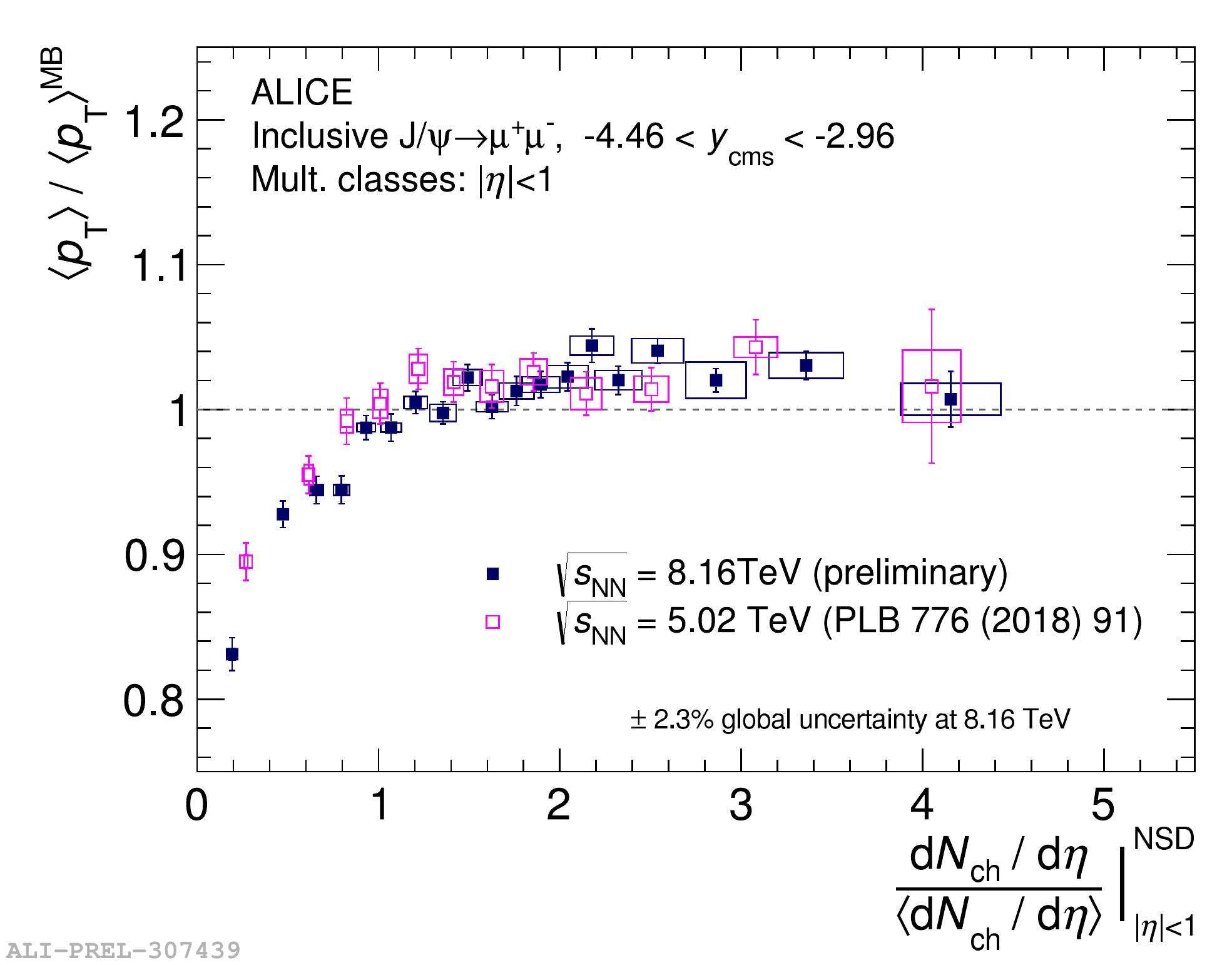}
%\end{floatrow}
%\caption{Relative yield and mean transverse momentum of $J/\psi$ as a function of relative charged-particle multiplicity for p \textendash Pb collisions at $\sqrt{s_{\rm{NN}}}$ = 8.16 TeV and the %results are compared with that of $\sqrt{s_{\rm{NN}}}$ = 5.02 TeV. The result for relative yield of inclusive $J/\psi$ production at forward ($2.03 < y_{\textrm{cms}} < 3.53$), backward ($-4.46 < %y_{\textrm{cms}} < -2.96$) are shown in left panel. The result for relative mean transverse momentum of $J/\psi$ at $-4.46 < y_{\textrm{cms}} < -2.96$ are shown in right panel.} 
\caption{Relative yields (left) and mean transverse momentum (right) of J$/\psi$ as a function of relative charged-particle multiplicity for p--Pb collisions at $\sqrt{s_{\rm{NN}}}$ = 8.16 TeV and $\sqrt{s_{\rm{NN}}}$ = 5.02 TeV. In the left panel the results are shown at both forward ($2.03 < y_{\rm{cms}} < 3.53$) and backward ($-4.46 < y_{\rm{cms}} < -2.96$) rapidities.  In the right panel only the backward rapidity result is shown, but similar trends are also observed at forward rapidity.} 
\label{fig2}
\end{center}
\end{figure*}

The multiplicity dependence of~J$/\psi$ production as a function of charged-particle multiplicity has also been studied in p--Pb collisions at $\sqrt{s_{\rm{NN}}}$ = 8.16 TeV.~The result is compared to that obtained in p--Pb collision at $\sqrt{s_{\rm{NN}}}$ = 5.02 TeV and is shown in the left panel of Fig.~\ref{fig2}. The measurements have been performed in different rapidity ranges by inverting the directions of the lead and proton beams. The backward rapidity result ($-4.46 < y_{\rm{cms}} < -2.96$) corresponds to J$/\psi$ produced in the direction of the Pb beam, whereas the forward rapidity result ($2.03 < y_{\rm{cms}} < 3.53$) corresponds to J$/\psi$ produced in the direction of the proton beam. We observe a colliding energy independent but rapidity dependent behavior for p--Pb collisions. A saturation towards higher multiplicities is observed at forward rapidity. In this kinematic region, the~proton~probes the small Bjorken-$x$ region of the Pb-nucleus, where cold nuclear matter effects, e.g. gluon saturation and/or gluon shadowing, are expected. 

Similarly, the J$/\psi$  $\langle {p_{\rm{T}}} \rangle$ has been measured as a function of the relative charged-particle density at $\sqrt{s_{\rm{NN}}}$ = 8.16 TeV for p--Pb collisions and compared to that measured at $\sqrt{s_{\rm{NN}}}$ = 5.02 TeV. We observed that the relative J$/\psi$  $\langle {p_{\rm{T}}} \rangle$ is independent of rapidity as well as center-of-mass energy within the current uncertainties. It first increases with multiplicity, and then saturates to a similar value for both rapidity ranges.

%$\langle {p_{\rm{T}}}^{J/\psi} \rangle$

\section{Summary} 
\label{sec:sum}   
In this contribution, recent ALICE measurements of the multiplicity dependence of quarkonium production have been presented for pp collisions at $\sqrt{s}$ = 5.02 and 13 TeV at forward rapidity. A similar study, performed in p--Pb collisions at $\sqrt{s_{\rm{NN}}}$ = 8.16, along with a measurement of the multiplicity dependence of the J$/\psi$ mean transverse momentum have also been presented. The comparison of these new results to that obtained at mid-rapidity hints for possible autocorrelations and jet bias effects when the J$/\psi$ and the charged-particle multiplicity are measured in the same rapidity range.

{}

\end{document}